\documentclass[aps,preprint,tightenlines]{revtex4}
\usepackage{amssymb}
\usepackage{epsfig}
\begin{document}

\title{Critical behavior of systems with long-range interaction in
restricted geometry}
\author{H. Chamati}
\email{chamati@issp.bas.bg}
\affiliation{Institute of Solid State Physics, Bulgarian Academy of
Sciences, 72 Tzarigradsko Chauss\'ee, 1784 Sofia, Bulgaria.}
\author{N.S. Tonchev}
\email{tonchev@issp.bas.bg}
\affiliation{Institute of Solid State Physics, Bulgarian Academy of
Sciences, 72 Tzarigradsko Chauss\'ee, 1784 Sofia, Bulgaria.}

\begin{center}
\underline{
\textsf{Accepted for publication in Modern Physics Letters B}
}
\end{center}

\begin{abstract}
The present review is devoted to the problems of finite-size
scaling due to the presence of long-range interaction decaying at
large distance as $1/r^{d+\sigma}$, $\sigma>0$. The attention is
focused mainly on the renormalization group results in the
framework of ${\cal O}(n)$ $\varphi^{4}$ - theory for systems with
fully finite (block) geometry under periodic boundary conditions.
Some bulk critical properties and Monte Carlo results also are
reviewed.  The role of the cutoff effects as well their relation
with those originating from the long-range interaction is also
discussed. Special attention is paid to the description of the
adequate mathematical technique that allows to treat the
long-range and short-range interactions on equal ground.
 The review closes with short discussion of some open
problems.
\end{abstract}

\maketitle

\section{Introduction}
Any system which has a finite size $L$ in at least one space
dimension we will call a finite-size system or a system with
restricted geometry. In such systems  singularities in the
thermodynamic functions at the critical point may occur only in
the thermodynamic (bulk) limit, taken in at least $d_{l}$ dimensions
($d_{l}$ is the lower critical  dimension). Mainly three specific
geometries (with periodic boundary condition imposed along the
finite dimensions) are of particular interest: i) the fully finite
cube $L^{d}$ , ii) the $d$-dimensional layer (or film)
$L^{1}\times \infty^{d-1}$  and iii) the infinitely long cylinder
$L^{d-1}\times\infty$. Further in this review we will not discuss
the latter two cases. So far, in the context of long-range (LR)
interaction they are studied only in the spherical $n=\infty$
limit. For the corresponding results one may consult the review
\cite{BT92} and the recent monograph \cite{brankov00}.

The geometry of real systems and Monte Carlo calculations usually
correspond to the fully finite size case. During the last two
decades the study of systems with restricted geometry has
undergone  an extensive development and gains still growing
importance for the theory of critical phenomena
\cite{brankov00,barber83,cardy88,privman90,krech94}. Generally
speaking the critical behavior depends  essentially on geometry,
boundary conditions,  and on the universality class of the bulk
system.

It is well known that the universality class  to which  the
critical behavior of a bulk system at a second order phase
transition belongs depends upon the dimensionality of the space
$d$, the number of the components of the order parameter $n$, the
symmetry of the Hamiltonian (either in spin-space or in coordinate
space) and the interaction potentials
\cite{aharony76,A78,zinnjustin96,cardy96}. One of the most
commonly studied interaction potentials, is the one corresponding
to LR ferromagnetic interaction decaying algebraically with the
spin separation $r$ as $r^{-d-\sigma}$. The parameter $\sigma>0$
controls the range of the interaction. The investigation of such
systems was initiated by Joyce in his paper on the phase
transition in  the ferromagnetic spherical model\cite{joyce66a}.
The interest in this type of interaction is tightly related to the
exploration of the critical behavior of systems with restricted
dimensionality in which no phase transition occurs otherwise (see
\cite{romano2000a,romano2000b,bruno2001} and references therein).
Here the condition $\sigma>0$ is needed to avoid an ill-defined
thermodynamic limit. In the limit $\sigma \rightarrow 0^{+}$ the
interaction (after appropriate renormalization) equals the
equivalent-neighbors interaction \cite{b90}.

The opposite case, corresponding to a LR interaction with $-d \leq
\sigma \leq 0$, sometimes called also nonintegrable interaction,
has been studied in connection with the so called "non-extensive
thermodynamics" (see \cite{Tsalis} and references therein). In
spite of the recent growing interest, the case of nonintegrable
interaction is beyond the scope of the present study.

The LR interaction, $r^{-d-\sigma}$, enters the expressions of the
theory only through its Fourier transform. The corresponding small
${\mathbf q}$ expansion of the Fourier transform has the general
form
\begin{equation}\label{1}
v({\mathbf q}) = v_{0} + v_{2}{\mathbf q}^{2} +v_{\sigma} {\mathbf
q}^{\sigma}+w({\mathbf q}) \qquad  0< \sigma \neq 2,
\end{equation}
with $w({\mathbf q})/{\mathbf q}^\sigma\to0$, for ${\mathbf q}\to
0$,  i.e in the long-wavelength approximation. Note that the case
$\sigma=2$ leads to logarithmic factors in (\ref{1}) which are not
allowed to enter the analysis below. However, further on we will
formally relate $\sigma=2$, to the
short-range (SR) interaction  since then~(\ref{1}) is the Fourier
transform of an interaction decaying exponentially with distance.
Depending on whether $\sigma \lessgtr 2$ in~(\ref{1}) we will
speak about {\it leading} or {\it subleading} LR term
respectively. Since the SR term $\backsim {\mathbf q}^{2}$ in
(\ref{1}) always exists the problem in the momentum space must be
considered with necessity as an interplay between SR and LR
effects.

This review is arranged as follows. In Section \ref{bulk} we
summarize the results on the critical behavior of bulk systems
with the LR interaction. Some general remarks on the finite-size
scaling (FSS) for such systems are presented in Section \ref{FSS}.
We review some results on the finite-size critical behavior of
systems with leading LR interaction in Section \ref{LLR} and for
the subleading one in Section \ref{SLR}. Section \ref{Binder}
specially deals with Binder's cumulant as an utilized tool to
compare the theory to data from Monte Carlo simulations. Some
remarks on the cutoff effects are devised in Section \ref{cutoff}.
In Section \ref{unsolved} we deliberately emphasize some unsolved
issues. In \ref{app}, we present some mathematical details in
calculating lattice sums in the case of LR interaction.

\section{Basic results in the bulk case}\label{bulk}
The results concerning the bulk critical behavior in the simplest
case of the spherical limit was analyzed in details in
\cite{joyce66b} (for a recent review, see Chapter 3 of
\cite{brankov00}). These results were generalized to the ${\cal
O}(n)$ vector $\varphi^4$ model by means of perturbation theory in
combination with the renormalization group (RG) technique near the
upper critical dimension
\cite{fisher72,sak73,Yam77,yamazaki77,GT,Hok} $d=2\sigma$, lower
critical dimension \cite{BZG,sak77,BCS82} $d=\sigma$, and the
$1/n$-expansion \cite{suzuki73,M73}. Computer simulations also
contributed to the exploration of the critical properties of such
systems \cite{LB97,romano96,GU88,luijten99,KL00,luijten2001,LB02}.

In the case of LR interaction, some RG predictions, e.g. using the
$\epsilon$-expansion, can be verified in an ideal testing ground
since the value of $\epsilon=2\sigma-d$ or $\epsilon=d-\sigma$
would be small enough for integer values of the dimensionality. In
this context the outcome of the computer simulations, obtained by
means of the Monte Carlo method, concerned mainly Ising systems
$(n=1)$ with classical (mean-field) critical behavior
\cite{LB97,luijten2001}, i.e. $0<\sigma<d/2$ with $d=1,2,3$, and
nonclassical critical behavior \cite{KL00,luijten2001,LB02}, i.e.
$d/2<\sigma$ with $d=1,2$. Some comparisons with  rigorous results
were obtained for low dimensional systems in
\cite{romano2000a,romano2000b} and for $d$-dimensional systems in
\cite{VE}.

It is worth mentioning that the critical behavior depends strongly
upon the interaction parameter $\sigma$. When $\sigma<2$ the
expansion ~(\ref{1}) has been used for detailed investigations of
the critical behavior of ${\cal O}(n)$ models including questions
like the $\sigma$, $d$ and $n$ dependence of the critical
exponents and critical amplitude ratios, as well as for
determining the universal scaling functions. In this case, the LR
term is {\it leading} and the critical exponents of the system are
$\sigma$ dependent. By increasing $\sigma$, a crossover from LR
critical behavior to SR one takes place. This issue has been
the matter of a long-standing debate in the
literature. As a measure for the crossover one may consider the
Fisher exponent $\eta$.

First, it has been argued, that in the interval $d/2<\sigma<2$ and
above the critical temperature $T_{c}$, the critical exponent
$\eta$ is equal to $2-\sigma$ with no corrections to order
$\epsilon^{2}$ and $\epsilon^{3}$ (at least), see\cite{fisher72},
and to ${\it O}(n^{-2})$, see \cite{suzuki73,M73}. It is
reasonable to believe that $\eta$ "sticks" to this value to all
order in $\epsilon$ \cite{aharony76}. So $\eta=\eta(\sigma)$ as a
function of $\sigma$, is not continuous at the  "crossover point"
$\sigma=2$, since the SR exponent $\eta(2) \neq 0$.

Later it was pointed out, in the cases of $(4-\epsilon)$
dimensions \cite{sak73} and in $(2+\epsilon)$ dimensions \cite{sak77}
that the crossover from LR to SR  critical behavior is shifted
and occurs at  a "critical" value of $\sigma$ given by
$\sigma_{c}=2-\eta(2)$. As a result $\eta=\eta(\sigma)$ is  a
continuous function in $\sigma$ at $\sigma=2$, since $\eta(\sigma
_{c})=\eta(2)$. Further support for this assertion was obtained in
the framework of different perturbation schemes in
\cite{Yam77,yamazaki77} and \cite{Hok}. In terms of the RG
language one can say that, by crossing the border  point $\sigma
_{c}$ the corresponding LR and SR fixed points exchange
stabilities and in a non vanishing range $2-\eta(2)<\sigma<2$ LR
perturbations are irrelevant (see also \cite{BCS82} where the
problem in the more general case of $(n,d,\sigma)$ space was
considered). This statement, which can be determined also from
more general considerations (see for example p.71 in
\cite{cardy96}) seems to be accepted in the literature. However,
some doubts arise since it conflicts with the opposite statement
of~\cite{VE} at least for the non Ising case $n\geq 2$ obtained
on a most rigorous level. Furthermore there is the criticism  in
\cite{GT} against the use of $(2-\sigma)$ as small perturbation
parameter. If one reconsiders the problem with $(2\sigma-d)$ as a
small parameter in conjunction with the use of the perturbation
scheme of \cite{GT}, previous results that LR fixed point is
stable up to $\sigma=2$, are restored. A recent attempt to
reconsider this issue using Monte Carlo simulations \cite{LB02}
shows unambiguously, a crossover at $\sigma_{c}$, in agreement
with \cite{sak73,Hok}, but there only the Ising  case ($n=1$) is
considered. Consequently, the situation seems to be settled only
for the case $n=1$.

When $\sigma>2$ one usually considers the model as equivalent to a
model with SR interaction, since it is widely accepted
\cite{fisher72,sak77,BCS82,suzuki73} that, the LR terms (i.e. the
third and etc. in Eq.(\ref{1})), do not contribute to the critical
behavior of the bulk system, consequently this term has been
always omitted in the computational analysis. Indeed, in this
case, the critical exponents does not depend on the parameter
$\sigma$.

\section{Some general remarks on the finite-size scaling}\label{FSS}
Scaling is a central idea in critical phenomena near a continuous
phase transition and in the field theory. In both cases the
singular behavior emerges from the overwhelming large number of
degrees of freedom, corresponding to the original cutoff scale,
which need to be integrated out leaving behind long wavelength
with smoothly varying momenta. This behavior is controlled by a
dynamically generated length scale: the bulk correlation length
$\xi$. Such a fundamental idea is difficult to test theoretically
because it requires the study of a huge number of interacting
degrees of freedom. Experimentally, however, one hopes to be able
to study {\it scaling in finite systems} near a second order phase
transition. Namely the system is confined to a finite geometry and
the FSS theory is expected to describe the behavior of the system
near the bulk critical temperature (for a review on the FSS theory
see \cite{brankov00,barber83,privman90}). In a few words the
\textit{standard} FSS is usually formulated in terms of {\it only
one} reference length - the bulk correlation length $\xi$. For a
system with finite linear size $L$, the main statements of the
theory are:

{\bf i)} The only relevant variable in terms of which the
properties of the finite system depend in the neighborhood of the
bulk critical temperature $T_c$ is $L/\xi$.

{\bf ii)} The rounding of the  the thermodynamic function
exhibiting singularities at the bulk phase transition in a given
finite system sets in when $L/\xi=O(1)$.

The tacit assumption is that all other reference lengths  are
irrelevant and will lead {\it only to corrections} towards the
above picture. Moreover, the crucial point in the  FSS theory is
that we always assume: the {\it finite size} $L$ of systems under
consideration and the correlation length $\xi$ are large in the
microscopic scale. This means $ L\gg a$ and  $\xi \gg a$, where
$a$ is the lattice spacing. The {\it scaling} limit of a quantity
is called its value when all corrections involving the ratios
$a/L$ and $a/\xi$ are neglected. One postulates  FSS limit as:
\begin{equation}
\frac{\xi}{a} \to \infty,\qquad \frac{L}{a} \to \infty, \qquad
\frac{L}{\xi} =const.
\end{equation}

In the literature there exist two conceptually different
approaches concerning the scaling regime. One inspired from the
condensed matter theory uses lattice language and models, in which
$a$ is fixed (while $L$ and $\xi$ are large, going to $\infty$).
The other one is inspired from  particle physics. It uses
continuum field theory models, when $L$ and $\xi$ are fixed (while
$a$ is infinitely small, going to zero). For studying a
dimensionless quantity when all the dependence on $a,L$ and $\xi$
is through the ratios $a/L$ and $a/\xi$  the two approaches seem
to be equivalent.

We shall be interested in the case where only
the bulk system displays  a phase transition. In such system there
is no fixed point for finite $L$ and so no crossover 
to any other fixed point as a function of $L/\xi$ can takes place. This
means that an $L$-independent RG procedure can be used. In the case
of LR interaction there are additional reasons (related with
possibilities of phase transitions in lowered dimensions)  to
study the case where finite system as well the bulk displays a
phase transition. However this more sophisticated case has been
studied only in the SR case \cite{CS02}.

So, we review the FSS properties of fully finite systems  in the
framework of  the continuous field theory. The effects of the
cutoff $\Lambda=\pi/a$ related with $a$ fixed will be discussed in
some details below.

In case of LR interaction there are two aspects of the theory
which should be mentioned here. First, in addition to the SR term
$|\mathbf{q}|^{2}$ in the propagators of the Feynman diagrams we
have also the LR term $|\mathbf{q}|^{\sigma}$. The last term
causes peculiar mathematical problems \cite{FP86,BT88,SP89}
concerning the evaluation of the lattice sums over
$\mathbf{q}$. Among the various methods  of accomplishing these
summation some details  of the method we prefer is presented in
\ref{app}, where its advantages are also discussed.
The second aspect arises specifically because the basic quantity
of the theory $\xi$ is unambiguously defined for SR interactions.
For the LR case it has to be defined in a more appropriate way
\cite{BD91}. The standard definitions of the bulk correlation length
are based on the asymptotic behavior of the pair correlation function
at large distance. The two most commonly used definitions of the bulk
correlation length are unambiguous in the case of exponential
decay of the bulk pair correlation function
$G_{\infty}(\mathbf{R};T)$ with the distance $R=|\mathbf{R}|$. One
defines
\begin{equation}
\xi_{1}(T)=-\lim_{R\rightarrow\infty}[R/\ln
G_{\infty}(\mathbf{R};T)].
\end{equation}
Alternatively, one may
consider the second moment of the bulk pair correlation function
and define the effective correlation radius,
\begin{equation}
\xi_{2}(T)=\left[\sum_{\mathbf{R}}R^{2}G_{\infty}(\mathbf{R};T)/G_{\infty}(\mathbf{R};T)
\right]^{1/2}.
\end{equation}
If the pair correlation function decays exponentially with
distance, the two definitions are equivalent. A distinctive
feature of the LR interaction is that the function
$G_{\infty}(\mathbf{R};T)$ decays as $R^{-d+\sigma}$ when
$R\rightarrow\infty$, for $T>T_{c}$ (see e.g.\cite{joyce66b}) and
both definitions yield $\xi_{1}(T)=\xi_{2}(T)=\infty $. One can
overcome this difficulty by following the approach proposed in
\cite{BT92} (see also p.138 of \cite{brankov00}) where instead of
$\xi$ some bulk \textit{characteristic length} $\lambda(T)$ is
used. This characteristic length determines the length-scale of
variation of the correlation function and diverges at the critical
point.

%\section{Basic results in the finite-size case}
As in the bulk limit ${\cal O}(n)$ models are the most often used
as laboratory tools on the basis of which one studies the scaling
properties of finite-size systems. The exactly investigated cases
are those corresponding to $n=1$, i.e. for the Ising model
(particular attention has been turned to the two dimensional
case), and the limit $n=\infty$, which includes the spherical
model (see e.g. Chapter 5 in \cite{brankov00} and p.141 in
\cite{privman90}{\bf )}. It is commonly assumed that the last one
is the only model which combines exact solubility even in the
presence of a magnetic field and direct relevance to the physical
reality. For that reason it is especially suitable for testing the
FSS hypotheses. For $n\ne1,\infty$ there are no exact results and
the commonly preferable analytical method for the derivation of
the properties of the corresponding models (like $XY$, i.e. $n=2$,
and Heisenberg, i.e. $n=3$) is that of the RG
theory. In addition an important amount of information for such
systems is derived by numerical simulations, normally via Monte
Carlo methods.

A systematic and controlled field-theoretical approach to the
quantitative computation of the thermodynamic moments was proposed
in the middle of 80's\cite{BZ85a,BZ85b} for studying FSS in the
case of SR interaction. It is based on the idea of using a mode
expansion, i.e. one treats the zero mode of the order parameter,
which is equivalent to the magnetization, separately from the
higher modes. The nonzero modes are traced over to yield an
effective Hamiltonian for the zero mode. This method is used in
combination with the loop expansion in the framework of the
minimal subtraction scheme. The method is quite general and was
proven to apply to a large extent to the investigation of FSS in
systems with LR interaction and with finite number of components
$n$ of the spin vector \cite{korucheva91}. However such
systems have been the object of analytical investigations mainly
recently \cite{luijten99,chamati01,chamati02,chamati2001}. For
example, it is possible to perform the quantitative computation of
the thermodynamic moments, usually used in numerical analysis.
These moments are related to the Binder's cumulant $B$ and to
various thermodynamic functions like the susceptibility.

The present wisdom is that in the finite size case the results
are quite different depending on whether the LR interaction is
leading or subleading. So, below we will consider separately the
cases of leading and subleading LR interaction.

\section{ Leading LR interaction}\label{LLR}
Recall that  the critical exponents depend upon the parameter
$\sigma<2$ controlling the interaction range. The FSS properties
(under periodic boundary conditions) in the spherical limit
are well established. It has been found that the
scaling properties of the system with SR interaction remains valid
also in the case of LR interaction (for a review see e.g Chapters
4 and 5 in \cite{brankov00} and references therein). For finite
$n$ a limited number of recent numerical results
\cite{romano96,luijten99,luijten2001}, as well as few analytical
works \cite{luijten99,luijten2001,korucheva91,chamati01,chamati02}
became available.

The nonlocal character of the LR interaction has been an obstacle
in investigating the critical behavior by means of numerical
methods. For that purpose different algorithms have been
developed. Using the Ewald method to evaluate the energy of a
given configuration, the critical behavior of such systems for
different values of $d$, $n$ and $\sigma$ has been
investigated\cite{romano96}. Another approach, analogous to FSS,
where the range of the interaction is cutoff at a certain value,
whence the name "finite-range scaling", is also used.\cite{GU88}
This method has been tested essentially on the one dimensional
Ising model in the classical (mean-field) as well as in the
nonclasical critical regimes. Results of the aforementioned
approaches have been judged comparable to the theoretical
predictions. These treatments required special efforts that
restrict the consideration to very small systems. Recently, the
problem was resolved by the use of cumulative probabilities within
the Wulff cluster algorithm \cite{LB97,luijten2001}. This method
has been applied  mainly to Ising systems in the mean-field
regime. The FSS hypotheses have been tested in the nonclassical
regime \cite{luijten99,luijten2001}. The common conclusion of all
these approaches is that in the case $d=\sigma$ such systems
exhibit a Kosterlitz-Thouless-like phase transition.

Analytically the nonclassical case has been studied first for
{$n=1$ }\cite{luijten99} and for $n\geqslant 1$
\cite{chamati01,chamati02}. It has been found that, as for the
bulk systems, the critical behavior depends on the small parameter
$\epsilon=2\sigma-d$. The scaling properties of the finite system
are not altered by the presence of the LR interaction and can be
written as
$$
{\cal X}=L^{-\gamma_x/\nu}{\cal F}(tL^{1/\nu}),
$$
where  $\gamma_x$ is the critical exponent of the observable
${\cal X}$ and $t=(T-T_{c})/T_{c}$. Some thermodynamic quantities,
like for instance, the shift of the critical temperature due to
the finite size effects, the susceptibility and the Binder's
cumulant at the bulk critical temperature $T_c$
\cite{luijten99,chamati01,chamati02} and above it,
\cite{chamati01,chamati02} as a function of $ \sqrt{\epsilon}$
have been obtained. The $d$ dependence of the finite size
properties has been considered in details in reference
\cite{chamati02} by means of a method, which consists of the use
of  the minimal subtraction scheme applied to a fixed space
dimensionality \cite{SD89}.

It has been shown that the critical behavior of the system is
dominated by its bulk critical behavior away from the critical
domain and that the FSS is relevant in the vicinity of the
critical point. A distinctive feature of the LR case is that for
$tL^{1/\nu}\gg1$ the finite-size corrections are not exponentially
small as in the SR case; they vary instead as power-in-L law,  in
both the spherical $n=\infty$ \cite{SP89,BD91} and $n\neq\infty$
cases \cite{chamati01,chamati02}.

In \cite{korucheva91} the behavior of the coupling constants
during the crossover from LR to SR interactions i.e. the limit
$\sigma\rightarrow2$, was considered. To one-loop order it has
been concluded that the renormalized values of the temperature and
the coupling constant are continuous functions of $\sigma$.

\section{Binder's cumulant}\label{Binder}
As it was mentioned above a quantity of central interest in the
study of the critical behavior of finite size systems  is the
Binder's cumulant ratio $B$. Here we present the analytical result
for $B$ ($B=1-M_{4}/3M_{2}^{2}$, where $M_{2n}$ is the $n$-th
moment; for some details, see\cite{BZ85a} and also
\cite{luijten99,chamati01}). Close to the critical point we have
\begin{eqnarray}\label{amplitude}
B&=&1-\frac {n}{12}\frac{\Gamma^2\left(\frac14n\right)}
{\Gamma^2\left(\frac14(n+2)\right)}\left\{1-
z\left[\frac{\Gamma\left(\frac14(n+6)\right)}
{\Gamma\left(\frac14(n+4)\right)}+
\frac{\Gamma\left(\frac14(n+2)\right)}
{\Gamma\left(\frac14n\right)}
-2\frac{\Gamma\left(\frac14(n+4)\right)}
{\Gamma\left(\frac14(n+2)\right)}\right]\right.\nonumber\\
&&\left.+z^2\left[\frac{\Gamma\left(\frac14(n+6)\right)
\Gamma\left(\frac14(n+2)\right)}
{\Gamma\left(\frac14(n+4)\right)\Gamma\left(\frac14n\right)}
+3\frac{\Gamma^2\left(\frac14(n+4)\right)}
{\Gamma^2\left(\frac14(n+2)\right)}-n-1\right] +
O\left(z^3\right)\right\}'.
\end{eqnarray}
At the fixed point  $z$ is given by \cite{chamati01}
($\epsilon=2\sigma-d$)
\begin{eqnarray}\label{zfixed}
z &=&\sqrt{\frac{n+8}{\epsilon}}\left[y-\frac\epsilon{2\sigma}y
\left(1-\frac{n-4}{n+8}\ln y\right)+2^{\sigma-1}\epsilon
\frac{n+2}{n+8}\Gamma(\sigma)F_{2\sigma,\sigma}\left(y\right)
\right.\nonumber\\
& &\left.-\epsilon2^{\sigma-2}y\Gamma(\sigma)
F_{2\sigma,\sigma}'\left(y\right)\right].
\end{eqnarray}
Here,
%Eq.~(\ref{zfixed})
 we have introduced the scaling variable
$y=tL^{1/\nu}$ and  the function
\begin{equation}\label{b2}
F_{d,\sigma}\left(y\right)=\int_0^\infty dx
x^{\frac\sigma2-1}E_{\frac\sigma2,\frac\sigma2}
\left(-\frac{yx^{\sigma/2}}{(2\pi)^\sigma}\right)
\left[\theta^d(x/\pi)-1-\left(\frac{\pi}{x}\right)^{d/2}\right].
\end{equation}
In Eq.(\ref{b2}) $E_{\alpha,\beta}(x)$ and $\theta(x/\pi)$ are the
generalized Mittag-Leffler and the reduced Jacobi $\theta_{3}$
functions, respectively (see also \ref{app} for details). The
function $F_{d,\sigma}\left(y\right)$ is well known
\cite{brankov00} in the theory of FSS.

Finally, let us notice that  one can easily see that the
expression (\ref{zfixed}) for $z$ as a function of $y$ verifies
the FSS hypotheses and, consequently, all the thermodynamic
functions, which indeed are $z$ dependent, do \cite{chamati01}. At
the critical temperature $T_c$ (i.e. $t=0$, and so $y=0$), we
obtain
\begin{equation}\label{z0}
z_0=\sqrt\epsilon\left[\frac{n+2}{\sqrt{n+8}}\sqrt{
\frac{\Gamma(\sigma)}{2\pi^\sigma}}F_{2\sigma,\sigma}(0)+
O(\epsilon)\right],
\end{equation}
where the coefficient $F_{2\sigma,\sigma}(0)$ appearing in the
right hand side of (\ref{z0}) can be evaluated \cite{CT00}
analytically as well as numerically for different values of the
interparticle interaction range $\sigma$
\begin{equation}\label{values}
F_{2\sigma,\sigma}\left(0\right)=\left\{
\begin{array}{ll}
2\zeta\left(1/2\right), &\sigma=1/2,\\[.3cm]
4\zeta\left(1/2\right)\beta\left(1/2\right), \ \ \ \ \ &\sigma=1,
\\[.3cm]
-4.82271993, &\sigma=3/2,\\[.3cm]
-8\ln2, &\sigma=2.
\end{array}
\right.
\end{equation}
Here $\zeta(x)$ is the Riemann zeta function with
$\zeta\left(\frac12\right)=-1.460354508 ...$ and $\beta(x)$ is the
analytic continuation of the Dirichlet series:
$$
\beta(x)=\sum_{\ell=0}^\infty\frac{(-1)^\ell}
{\left(2\ell+1\right)^{x}},
$$
with $\beta\left(\frac12\right)=0.667691457 ...$. Note that the
function $F_{2\sigma,\sigma}(0)$ increases as the parameter
$\sigma$ vanishes.

\begin{figure}
\epsfxsize=4in
\centerline{\epsffile{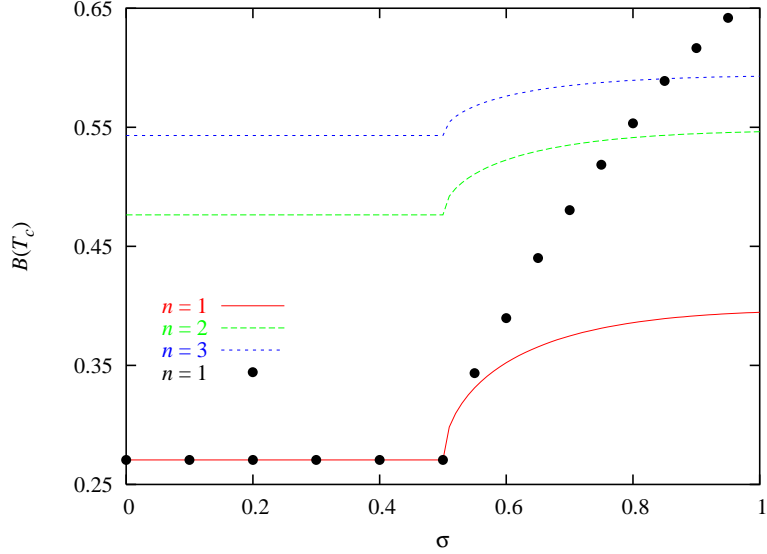}}
\caption{The cumulant ratio $B$ as a function of the interaction
range $\sigma$ for $d=1$. The case $0<\sigma<d/2$ corresponds to
classical (mean-field) regime.The Monte Carlo results "$\bullet$" follow
from \protect\cite{LB97} for $0<\sigma\leq0.5$ and from
\protect\cite{luijten99} for $0.5<\sigma\leq1$.} \label{fig2}
\end{figure}

\begin{figure}
\epsfxsize=4in
\centerline{\epsffile{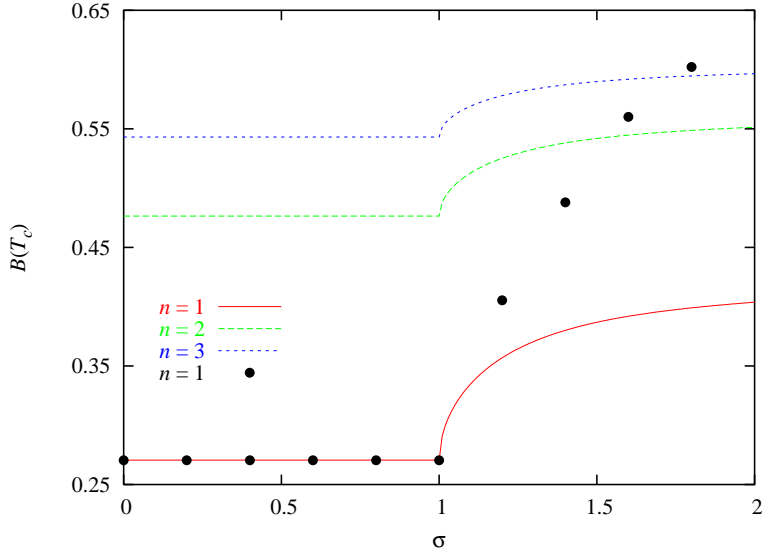}}
%\vspace{0.1in}
\caption{The same as Fig. \protect\ref{fig2} for
$d=2$}\label{fig3}
\end{figure}

Numerical values for the Binder's cumulant ratio (\ref{amplitude})
can be obtained by replacing the value of $z_0$ form (\ref{z0})
and taking some specific values of the small parameter $\epsilon$.
The behavior of the universal constant $B$ for the cases $d=1,2$
is presented in Figures \ref{fig2} and \ref{fig3}, respectively.
Note that the scaling variable $z$ is proportional to
$\sqrt\epsilon$ as it was found previously (see \cite{BZ85a} for
example) in the case of SR forces. Furthermore for $n=1$ it is in
full agrement with the result  obtained in reference
\cite{luijten99} devoted to the exploration
of FSS in ${\cal O}(n)$ systems with LR
interaction. Notice that in \cite{luijten99} (see also
\cite{luijten2001}) the mathematically ensuing  pertinent
integrals have to be evaluated only numerically, due to the choice
of a parametrization, e.g. (\ref{Scr}), that does not reduce the
$d$-dimensional problem to the effective one-dimensional one. The
approach proposed in \cite{chamati01} based on the parametrization
(\ref{BTr}) is more efficient in the sense that the corresponding
final expressions can be handled by analytical means \cite{CT00}.

It has been found \cite{luijten99}, using the Monte Carlo method,
that the amplitude ratio $Q=M_{2}^{2}/M_{4}$ (which is related to
$B$) is a linear function of $\epsilon$ (see also Fig.1 and Fig.2),
while the analytical result (\ref{z0}) shows an expansion in powers of
$\sqrt\epsilon$ \cite{luijten99,chamati01}. It is possible that 
higher orders in $\epsilon$ could improve the result. Better agreement 
could also be obtained by using the method developed in \cite{chamati02}.

\section{ Subleading LR interaction}\label{SLR}
Let us first note that the most prominent example for such kind
of interaction in the case $d=3$ is the van der Waals interaction 
$1/r^{6}$ for which we have $\sigma=3$. One can easily see that in 
this case the Fourier transform is indeed of the type (\ref{1}).

The fact that the critical behavior is not affected by the LR
interaction if $\sigma>2$ is well established in the bulk case.
Analytically, in the framework of $n=\infty$ model it has been
shown \cite{dantchev01} that such a statement is incorrect for
finite-size systems because of finite-size contributions due to
the {\it subleading}, $\sigma>2$, term in the interaction. It has
been shown also that the same remains true for a finite number of
component by means of the RG techniques \cite{chamati2001}. The
authors of \cite{chamati2001} have investigated the FSS behavior
of a fully finite ${\cal O}(n)$ system with periodic boundary
conditions and in the presence of a LR interaction that {\it does
not alter the SR exponents} of its critical behavior. The small
$|{\mathbf q}|$ expansion of the Fourier transform of the
interaction $v(\mathbf q)$ is supposed to be of the form
~(\ref{1}) with $2<\sigma<4$. For such a system, it has been
demonstrated that all the thermodynamic functions can be expressed
in a scaling form as
$$
{\cal X}=L^{-\gamma_x/\nu}{\cal F}(tL^{1/\nu},bL^{2-\sigma-\eta}),
$$
where $b$ is a model (nonuniversal) constant. Note that one needs
two scaling variables  in order to describe in a proper way the
finite size behavior of these quantities.

\begin{figure}
\epsfxsize=4in
\centerline{\epsffile{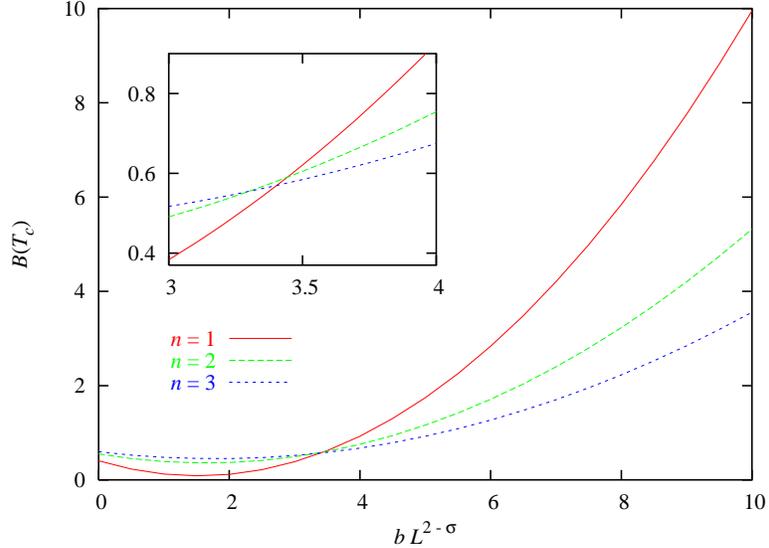}}
\caption{Binder's cumulant ratio az a function of the variable
$bL^{2-\sigma}$}\label{fig4}
\end{figure}

At the critical point the Binder's cumulant is given
\cite{chamati2001} by (\ref{amplitude}) with
\begin{equation}\label{fpsl}
z_{0}=-\sqrt\epsilon\frac{\sqrt{32}}{\pi}\frac{n+2} {\sqrt{n+{\bf
8}}}\left[\ln2+b\varpi_{\sigma}L^{2-\sigma}+ O(\epsilon)\right],
\end{equation}
where $\varpi_{\sigma}=(2\pi)^{\sigma-2}(1-4^{\sigma/2-1})
\zeta(1-\sigma/2)\zeta(2-\sigma/2)$. So, in this case $B$ is not
an universal constant. The behavior of $B$ as a function of system
size $L$ is presented in Fig. \ref{fig4}. Equation (\ref{fpsl}) is
a generalization for the case under consideration of the result
obtained in \cite{BZ85a} for SR interaction.

When $tL^{1/\nu}\gg1$, it was found that the susceptibility
approaches its bulk value not in an exponential-in-$L$, as it was
commonly believed to be the case for systems with short-range
critical exponents, but in a power-in-$L$ way of the order of
$bL^{-(d+\sigma)}$. The last goes beyond the standard formulation
of the finite-size scaling, but is completely consistent with the
intrinsic large-distance power-law behavior of the correlations in
systems with subleading LR interactions. In the spherical limit
such large-distance power-law behavior is shown exactly in
\cite{D01}.

For an \textit{arbitrary} value of $\sigma$, away from the critical 
point, i.e. in the region $tL^{1/\nu}\gg1$, the Binder's cumulant 
ratio is given by $B=1-\frac13\left(1+\frac2n\right)$, with finite 
size correction falling off in a power law. This result corresponds 
to a $n$-dimensional Gaussian distribution for $n$ independent
components of the vector variable. Obviously, all the values lie
in the interval from $ B=0$ (Ising model, $n=1$) to $ B=2/3$
(spherical model, $n=\infty$).

\section{Lattice and continuum models and the effect of
cutoff}\label{cutoff}
In the above, we have concentrated our attention on how to
separate the size dependence in the continuum (scaling) limit,
where the linear size as well as the correlation length tend to
infinity, but keeping their ratio a finite quantity. In this
theory the cutoff is sent to infinity and the lattice spacing
completely disappears. This is precisely the regime in which one
expects FSS to hold \cite{BZ85a,BZ85b} and this statement, as it
was pointed out, does not depend on the range of the interaction.

In order to clarify the effect of a (sharp) cutoff it is
sufficient to consider the pure SR case. The finite cutoff effects
violate the FSS, see \cite{chen99a,chen99b} and references
therein. The result is a system with finite-size behavior with
leading nonscaling term going as an inverse power law in $L$ that
depends also on $L\Lambda$. In \cite{chen99a} this is shown to be
tightly related to lattice effects in the system in
\textit{conjunction} with the long wavelength approximation.
Indeed this is not in conflict with the arguments proposed in
\cite{BZ85a,BZ85b} where the limit of infinite cutoff is
considered. Later, in \cite{dantchev01} one argues : i) the
violation of scaling is related with the second term in
(\ref{ae1}) which in the case $a\neq0$ can not be omitted in the
region $L/\xi\gg1$, ii) the effect of the sharp cutoff is similar
to that found if one considers the effect of subleading LR
interaction. Obviously, this effective LR nature of the
finite-size behavior should  generate difficulties in the analysis
of the Monte Carlo data from simulations not only for SR, but for
LR systems as well. The last follows directly from (\ref{int1}).
Recently, the reconsideration of the problem in \cite{DKD} has
shown that the source of these obstacles is the artificial
cusp-like singularity
\begin{equation}\label{UDKD}
\left.\frac{\partial v({\mathbf q})}{\partial q_{i}}\right
\vert_{q_{i}=\Lambda} \not=\left. \frac{\partial v({\mathbf
q})}{\partial q_{i}}\right \vert_{q_{i}=-\Lambda}, \qquad
i=1,\ldots,d
\end{equation}
at the border of the Brillouin zone ($\Lambda\sim 1/a$). As a
result, if one treats properly the effects generated by the
momentum cutoff in the Fourier transform of the interaction
potential both, the lattice and continuum models may produce
results in mutual agreement independent of the cutoff scheme
\cite{DKD}.

\section{ Unsolved problems}\label{unsolved}
First, an area where the bulk RG predictions \cite{sak73,Hok} must
be tested by means of numerical methods is the case of $n>1$.  The
strong indication \cite{VE} that exactly here  some problems
exist, remains still valid. The numerical results of \cite{LB02}
shed some light on the Ising case $n=1$ but do not solve  the
longstanding controversy about the boundary between LR and SR
critical behaviors as a function of $\sigma$.

In this review the FSS properties of systems with
no crossover from the bulk fixed point to any other as a function 
of $L/\xi$ are considered. Let us note that within the SR models
this issue is studied in \cite{CS02}. The consideration of this 
crossover in the LR is missing. On the other hand, e.g. systems 
with slab and cylinder geometries  and $n\neq \infty$ have been 
studied only in the SR case. The first one is related to the
statistical-mechanical Casimir effect in fluctuating systems
\cite{FG78}  and recently there has been an upsurge in theoretical
investigations\cite{brankov00,krech94,chen99b,DKD}. The second one
deals with the so called transfer matrix (or Hamiltonian)
formulation of the problem \cite{BZ85a} (see also Chapter 36.4 of
\cite{zinnjustin96}). A study of both cases in the context of the
LR interactions is a provocation for the theory and is of
undoubted experimental interest.

Some problems exist with the comparison of the analytical results
and the numerical data. It has been shown\cite{luijten99} (see
also \cite{luijten2001}) using Monte Carlo simulation (for $n=1$),
that the Binder's cumulant ratio is linear in $\epsilon$. The
analytical evaluation (\ref{amplitude}) for the ${\cal O}(n)$
symmetric $\varphi^4$ model, however, showed that it is linear in
$\sqrt\epsilon$. A possible way to resolve this controversy
between the Monte Carlo method and the analytical results is to
carry on finite-size calculations to higher loop
order\cite{luijten99}. This could ameliorate the analytical
results, which would be comparable to those obtained by numerical
simulations. However we would like to mention that higher loop
corrections that are dealt with through the minimal subtraction
scheme and the $\epsilon$-expansion  are not done, even for the
more simple case of SR interaction. We have witnessed that
two-loop calculations can be applied in many other investigations
like the problems taking into account disorder effects in finite
size systems \cite{chamati2001d}. Here, if $n=1$, the one-loop
fixed point is degenerate and  the first interesting results can
be obtained only if one considers two loops.

As it is discussed in the previous section in the long wavelength
approximation (\ref{1}), if one goes beyond the continuum field
theory, the lattice effects in conjunction with the sharp cutoff
generate nonuniversal finite-size terms. The suggestion that such
effects are artificial in the case of SR interaction \cite{DKD}
seems to be true also in the LR case. How to avoid the problem in
the framework of a concrete cutoff  scheme in the case of LR
interaction is still an open problem.

In recent years considerable attention has been paid to the
critical dynamics of systems with LR interaction \cite{CG00}.
It would be interesting to extend the corresponding results to
the finite-size case. Let us note that for the case of the SR 
interaction the theory of finite size effects in critical dynamics 
was developed in \cite{NZ87}(see also Chapter 36.6 
of \cite{zinnjustin96}).

The present review is devoted to the classical critical phenomena.
However, another interesting field is closely related to the
extensively investigated field of quantum critical points, i.e.
phase transitions occurring at zero-temperature \cite{S99}. Let us
just recall that in systems showing quantum critical behavior the
temperature plays two different roles. For temperatures low
enough, quantum effects are essential. In this case the
temperature affects the geometry to which the system is confined
adding a ``new'' size to the Eucludean space-time coordinate
system. By raising the temperature, the system is driven away from
the quantum criticality. At high temperatures, however, the size
in the ``imaginary-time'' direction becomes irrelevant in
comparison with all length scales in the system. In this case we
have a classical system in $d$ dimensions and the temperature is
just a coupling constant in the classical critical behavior. In
this context we find it useful to investigate the quantum models,
with LR interaction, e.g. considered in \cite{CT2000,CDT00} in the
large $n$ limit or in \cite{AB01} in the mean field and one-loop
RG theory.

\appendix
\section{The origin of the mathematical difficulties in evaluating the
lattice sums}\label{app} In order to introduce the reader to the
problems let us first consider in some details the SR case on a
$d$-dimensional hypercubic lattice $\mathbb Z^{d}$ with $N=N_{0}^{d}$ 
sites and periodic boundary conditions. The sites are given by
\begin{equation}
{\mathbf r}=a(n_{1},\ldots,n_{d}),
\end{equation}
where $n_{i}$ ranges over all distinct integer values
${\mathrm mod} N_{0}$, for $i=1,\ldots,d$.

Let us consider the dimensionless expression , entering most of
the mathematical analysis of the scaling properties of finite
systems
\begin{eqnarray}\label{ole}
&G&(N_{0},a|d)= \frac{m^{2-d}}{
(aN_{0})^{d}}\sum_{n_{1}=-N_{0}/2}^{N_{0}/2-1}\ldots
\sum_{n_{d}=-N_{0}/2}^{N_{0}/2-1}\frac{1}{m^{2}+
|\mathbf{q}|^{2}}\nonumber \\
&=& \frac{1}{ (amN_{0})^{d}}\int_{0}^{\infty}\exp(-t)
\left\{\sum_{n=-N_{0}/2}^{N_{0}/2-1} \exp \left[-\left(\frac{2\pi
n}{amN_{0}}\right)^2t\right]\right\}^{d}dt,
\end{eqnarray}
where $q_{i}=\frac{2\pi n_{i}}{aN_{0}}$ and $N_{0}$ is even. For
the sum
\begin{equation}
Q_{N_{0}}(t)=\left(\frac{1}{amN_{0}}\right)\sum_{n=-N_{0}/2}^{N_{0}/2
- 1} \exp \left[-\left(\frac{2\pi n}{amN_{0}}\right)^2t\right],
\end{equation}
using the result of \cite{dantchev01}, we have
\begin{eqnarray}\label{ae1}
Q_{N_{0}}(t)&\cong& \frac{1}{\sqrt{4\pi t}}
\left[\mathrm{erf}\left(\frac{\pi t^{1/2}}{am}\right) \right]
-\frac{2\pi^{2}t}{3}\frac{1}{am}\exp\left[-\pi^{2}t\left(\frac{1}{am}
\right)^{2}\right] \nonumber\\
&+& \frac{1}{\sqrt{\pi t}} \left\{ \sum_{l=1}^{\infty}
\exp[-l^{2}(amN_{0})^{2}/4t]\right\},
\end{eqnarray}
valid in the large $N_{0}$ asymptotic regime. The first and the
second terms in the above equation are size independent and are
precisely the infinite volume limit of $Q_{N_{0}}(t)$. The
ultraviolet divergences which may appear in the theory when the
lattice spacing vanishes are related with the first term, for
which one must perform the required subtractions (see, e.g. p. 177
of \cite{zinnjustin96}). The second term depends on the ratio
$\xi/a\equiv1/ma$ and so if $a \to 0$ or $\xi \to \infty$ it is of
order
\begin{equation}\label{cae}
O\left((\xi/a) t e^{-\pi^{2}t(\xi/a)^{2}}\right).
\end{equation}
In the continuum limit such terms are  exponentially small and
must be omitted.

The different finite-size regimes are governed by the ratio
$\xi/L$  in the third term. They are independent of the
microscopic details, e.g. lattice spacing $a$. When $\xi \gg a$,
the {\it universal properties} can be described by a continuous
field theory. Let us  recall that the finite linear dimension
$L=N_{0}a$, in the case of continuous finite volume means that $a
\to 0$ and simultaneously $N_{0}\to \infty$.

In the continuum limit, since $\mathrm{erf}(\pi t^{1/2}/{\it a
m})=1$, we end up with the result
\begin{equation}
G(N_{0},0|d)=G(\infty,0|d)|_{L=\infty} + g(L|d),
\end{equation}
where the size dependence is contained in the function
\begin{eqnarray}
g(L|d)&=& \frac{1}{\pi^{d/2}}\int_{0}^{\infty}
\frac{\exp(-t)}{t^{d/2}} \sum_{{\mathbf l}(d)=1}^{\infty}
\exp\left(-m^{2}|{\mathbf l}(d)|^{2}L^{2}/4t\right)dt \nonumber \\
&=&\frac{1}{(4\pi)^{d/2}}\int_{0}^{\infty}
\frac{\exp(-t)}{t^{d/2}}\left\{\left[\theta\left(\frac{m^{2}L^{2}}{4\pi
t} \right)\right]^{d} -1 \right\}dt,
\end{eqnarray}
and where
\begin{equation}
\theta(x)=\sum_{n=-\infty}^{\infty}\exp(-\pi x n^{2}).
\end{equation}
The bulk part $G(\infty,0|d)|_{L=\infty}$ contains poles at
$d=2,4,\ldots$. The finite size correction $g(L|d)$ combined with
the corresponding contribution from the counterterm yields
\cite{BZ85a,BZ85b} (to one - loop order) a finite-size shift of
the "mass term" $m^{2}$.

In order to investigate the FSS properties of systems with LR
interaction, one can use a suitable mathematical method allowing
to simplify the analytical calculations. In the above case of SR
interaction it has been possible to replace the summand in
eq.~(\ref{ole}) by its Laplace transform
\begin{equation}\label{Scr}
\sum_{\mathbf{q}}\frac{1}{m^{2}+|\mathbf{q}|^{2}}=
\int_{0}^{\infty}dt\exp(-m^{2}t)
\left[\sum_{q}\exp(-q^{2}t)\right]^{d},
\end{equation}
where $\mathbf{q}$ and $q$ are d-dimensional and one-dimensional
discrete vectors, respectively. This is the so called Schwinger
parametric representation. The aim of this replacement is two
fold: i) to reduce the $d$-dimensional sum to  the corresponding
effective one-dimensional one, and ii) to give to the
dimensionality $d$ the status of a continuous variable. In the
case with the $\sigma$ term leading in (\ref{1}), one cannot just
use the Schwinger representation in its familiar form.
In\cite{BT88} the following generalization of (\ref{Scr}) has been
suggested
\begin{equation}\label{BTr}
\sum_{\mathbf{q}}\frac{1}{m^{2}+|\mathbf{q}|^{\sigma}}=
m^{\frac{4-2\sigma}{\sigma}} \int_{0}^{\infty}dt
Q_{\sigma}(m^{4/\sigma}t)\left[\sum_{q}\exp(-q^{2}t)\right]^{d},
\end{equation}
where the function $Q_{\sigma}(t)$ for $0<\sigma<2$ is given by
\begin{equation}\label{F}
Q_{\sigma}(t)=
\int_{0}^{\infty}dy\exp(-ty){\tilde Q}_{\sigma}(y),
\end{equation}
and
\begin{equation}\label{kernel}
{\tilde
Q}_{\sigma}(y)=\frac{1}{\pi}\frac{\sin(\sigma\pi/2)y^{\sigma/2}}
{1+2y^{\sigma/2}\cos(\sigma\pi/2)+y^{\sigma}}.
\end{equation}
From (\ref{BTr}) and (\ref{F}) for the summand in ~(\ref{BTr}) one
obtains
\begin{equation}\label{int1}
\frac{1}{m^{2}+|\mathbf{q}|^{\sigma}}= \int_{0}^{\infty}dt{\tilde
Q}_{\sigma}\left(\frac{t}{m^{\frac{4-2\sigma}{\sigma}}}\right)
\frac{1}{tm^{2}+|\mathbf{q}|^{2}}.
\end{equation}

\begin{figure}
\epsfxsize=4in \centerline{\epsffile{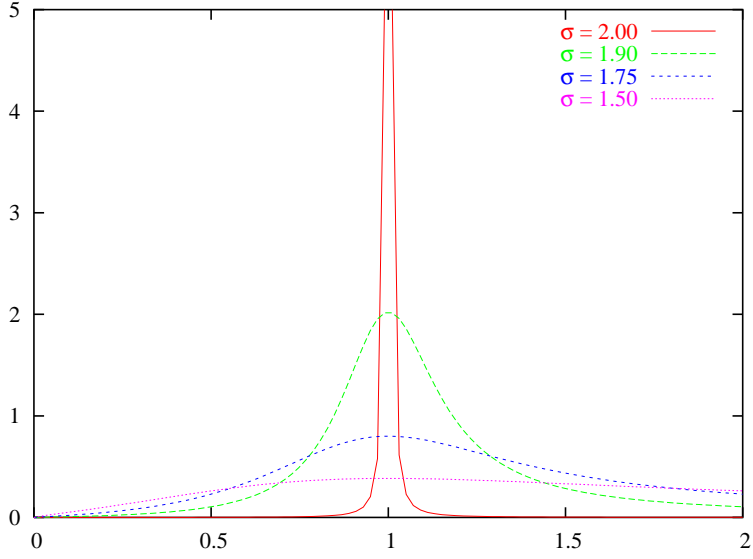}}
\caption{The dependence of kernel (\ref{kernel}) as a function of
$x$ for different interparticle interaction range
$\sigma=1.50;1.75;1.90$ and $2,00$.} \label{fig1}
\end{figure}

The integral representation (\ref{int1}) illustrates  the
mathematical difficulty  appearing in the LR case. It is shown
that by an additional integration the problem can be effectively
reduced to the SR case\cite{BT88}. The behavior of ${\tilde
Q}_{\sigma}(x)$ as a function of $\sigma$ (see Fig. \ref{fig1})
shows another aspect of the nonlocal character of the LR
interaction $1/r^{d+\sigma}$.

The function $Q_{\sigma}(x)$ is related \cite{B89} to the entire
function of the Mittag-Leffler type defined by the power series
\begin{equation}
E_{\alpha,\beta}(z)=\sum_{k=0}^{\infty}\frac{z^{k}}{\Gamma(\alpha
k+\beta)} \qquad \alpha>0.
\end{equation}
For a more recent review on these functions see\cite{GM97}. It has
been {demonstrated} \cite{B89} that
\begin{equation}
Q_{\sigma}(x)=x^{\sigma/2 -
1}E_{\sigma/2,\sigma/2}(-x^{\sigma/2}).
\end{equation}
and all the specific features of the finite size properties
of systems with LR interaction are a result of the analytical
properties of the  Mittag-Leffler type functions (for some
examples, see e.g. Chapters 5 and 6 of \cite{brankov00}).

In the case with both SR and LR terms the corresponding
expressions are significantly more complicated. When the LR term
is of the  peculiar type $\sigma=2-2\alpha$ (and the parameter
$\alpha\rightarrow0^{+}$) considered in the context of the bulk
crossover the following replacement of the SR propagator takes
place \cite{Hok}
\begin{equation}\label{lrt}
\frac{1}{m^{2}+|\mathbf{q}|^{2}}\rightarrow\sum_{l=0}^{\infty}
\frac{(-v_{\sigma})^{l}|\mathbf{q}|^{2l(1-\alpha)}}{(m^{2}+
|\mathbf{q}|^{2})^{1+l}}.
\end{equation}
In order to reduce the problem of evaluating the asymptotic
behavior of the sum over $\mathbf q$ to the corresponding one-dimensional
sum in the right hand side of (\ref{lrt}), the following identity
has been used\cite{korucheva91}
\begin{equation}\label{KTI}
\frac{|\mathbf{q}|^{2l(1-\alpha)}}{(m^{2}+|\mathbf{q}|^{2})^{1+l}}=
\frac{1}{\Gamma(1+l\alpha)}\int_{0}^{\infty}dx x^{l\alpha}
\quad_{1}F_{1}(1+l;1+l\alpha;-m^{2})\exp(-|\mathbf{q}|^{2}x).
\end{equation}
 Here, $\quad_{1}F_{1}$ is the degenerate hypergeometric function
(Kumer's function).  The same identity ~(\ref{KTI}) has been used
in \cite{dantchev01} to analyze the finite-size behavior of a
propagator with SR and subleading, i.e. $\sigma>2$, LR interaction
(treated as a perturbation) where terms similar to the r.h.s. of
~(\ref{lrt}) also appear. In the last case an expression for the
entire propagator (i.e. SR and LR interactions treated on equal
ground) can be obtained via contour integration on the complex
plane\cite{dantchev01}.

The identities (\ref{BTr}) and (\ref{KTI}) demonstrate that LR
case can be effectively reduced to the SR case with an integration
over an additional parameter. So, all the mathematical  machinery
developed for the SR case may be used without further
complications.

\begin{acknowledgments}
The authors thank J.G.Brankov  for reading the manuscript and
for his important comments. N.S.T. is also indebted to  D.M.Danchev for
acquainting with the article \cite{DKD} prior to publication and
for illuminating discussion on the cutoff effects.
\end{acknowledgments}

\end{document}